\documentclass[twoside,twocolumn,9pt]{article}

\usepackage{extsizes}
\usepackage[super,sort&compress,comma]{natbib} 
\usepackage[version=3]{mhchem}
\usepackage[left=1.5cm, right=1.5cm, top=1.785cm, bottom=2.0cm]{geometry}
\usepackage{balance}
\usepackage{mathptmx}
\usepackage{sectsty}
\usepackage{graphicx} 
\usepackage{lastpage}
\usepackage[format=plain,justification=justified,singlelinecheck=false,font={stretch=1.125,small,sf},labelfont=bf,labelsep=space]{caption}
\usepackage{float}
\usepackage{fancyhdr}
\usepackage{fnpos}
\usepackage[english]{babel}
\addto{\captionsenglish}{%
  
}
\usepackage{array}
\usepackage{droidsans}
\usepackage{charter}
\usepackage[T1]{fontenc}
\usepackage[usenames,dvipsnames]{xcolor}
\usepackage{setspace}
\usepackage[compact]{titlesec}
\usepackage{hyperref}
\usepackage{booktabs}
\usepackage{multirow}

\usepackage{epstopdf}

\definecolor{cream}{RGB}{222,217,201}

\begin{document}

\pagestyle{fancy}
\thispagestyle{plain}
\fancypagestyle{plain}{
\renewcommand{\headrulewidth}{0pt}
}

\makeFNbottom
\makeatletter
\renewcommand\LARGE{\@setfontsize\LARGE{15pt}{17}}
\renewcommand\Large{\@setfontsize\Large{12pt}{14}}
\renewcommand\large{\@setfontsize\large{10pt}{12}}
\renewcommand\footnotesize{\@setfontsize\footnotesize{7pt}{10}}
\makeatother

\renewcommand{\thefootnote}{\fnsymbol{footnote}}
\renewcommand\footnoterule{\vspace*{1pt}%
\color{cream}\hrule width 3.5in height 0.4pt \color{black}\vspace*{5pt}} 
\setcounter{secnumdepth}{5}

\makeatletter 
\renewcommand\@biblabel[1]{#1}            
\renewcommand\@makefntext[1]%
{\noindent\makebox[0pt][r]{\@thefnmark\,}#1}
\makeatother 
\renewcommand{\figurename}{\small{Fig.}~}
\sectionfont{\sffamily\Large}
\subsectionfont{\normalsize}
\subsubsectionfont{\bf}
\setstretch{1.125} 
\setlength{\skip\footins}{0.8cm}
\setlength{\footnotesep}{0.25cm}
\setlength{\jot}{10pt}
\titlespacing*{\section}{0pt}{4pt}{4pt}
\titlespacing*{\subsection}{0pt}{15pt}{1pt}

\fancyfoot{}
\fancyfoot[LO,RE]{\vspace{-7.1pt}\includegraphics[height=9pt]{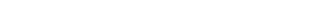}}
\fancyfoot[CO]{\vspace{-7.1pt}\hspace{13.2cm}\includegraphics{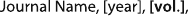}}
\fancyfoot[CE]{\vspace{-7.2pt}\hspace{-14.2cm}\includegraphics{head_foot/RF}}
\fancyfoot[RO]{\footnotesize{\sffamily{1--\pageref{LastPage} ~\textbar  \hspace{2pt}\thepage}}}
\fancyfoot[LE]{\footnotesize{\sffamily{\thepage~\textbar\hspace{3.45cm} 1--\pageref{LastPage}}}}
\fancyhead{}
\renewcommand{\headrulewidth}{0pt} 
\renewcommand{\footrulewidth}{0pt}
\setlength{\arrayrulewidth}{1pt}
\setlength{\columnsep}{6.5mm}
\setlength\bibsep{1pt}

\makeatletter 
\newlength{\figrulesep} 
\setlength{\figrulesep}{0.5\textfloatsep} 

\newcommand{\topfigrule}{\vspace*{-1pt}%
\noindent{\color{cream}\rule[-\figrulesep]{\columnwidth}{1.5pt}} }

\newcommand{\botfigrule}{\vspace*{-2pt}%
\noindent{\color{cream}\rule[\figrulesep]{\columnwidth}{1.5pt}} }

\newcommand{\dblfigrule}{\vspace*{-1pt}%
\noindent{\color{cream}\rule[-\figrulesep]{\textwidth}{1.5pt}} }

\makeatother

\twocolumn[
  \begin{@twocolumnfalse}
{\includegraphics[height=30pt]{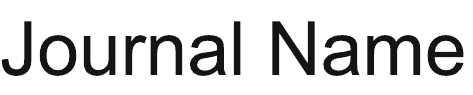}\hfill\raisebox{0pt}[0pt][0pt]{\includegraphics[height=55pt]{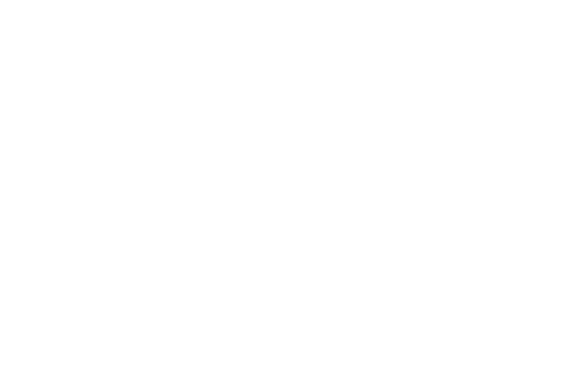}}\\[1ex]
\includegraphics[width=18.5cm]{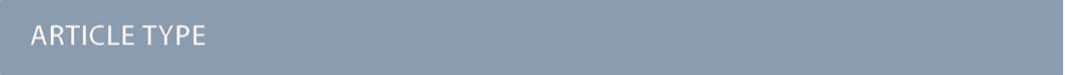}}\par
\vspace{1em}
\sffamily
\begin{tabular}{m{4.5cm} p{13.5cm} }

\includegraphics{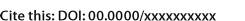} & \noindent\LARGE{\textbf{First-Principles Investigation of Multimodal Toxic Gas Sensing in Carbon-Tuned hBN–Graphene Alloys: Chemiresistive, Work-Function, and Optical Responses$^\dag$}} \\

\vspace{0.3cm} & \vspace{0.3cm} \\

 & \noindent\large{Tanjuma Shikder Jhumu\textit{$^{a}$} and  Ahmed Zubair\textit{$^{a,*}$}} \\

\includegraphics{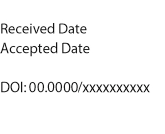} & \noindent\normalsize{Compact, reliable, and cost-effective gas sensors have become a highly demanding subject for safety management in the medical sector, chemical manufacturing, food quality monitoring, agriculture, and industrial safety. Hazardous gas emissions need to be controlled and monitored with fast-responsive and highly sensitive sensing devices. A first-principles study employing density functional theory (DFT) was used to investigate the adsorption behavior of Cl$_2$, CO, CO$_2$, NO, NO$_2$, and HCN gas molecules with our proposed alloys, which consisted of hexagonal boron nitride (hBN) and graphene (Gr). The alloy consisting of 22\% carbon (BNGr-2) was found to be most competent for sensing Cl$_2$, CO, CO$_2$, and HCN with sufficient adsorption energy, charge transfer, and bandgap alteration. However, NO and NO$_2$ gas molecules showed more engagement with 33\% carbon-proportioned alloy (BNGr-3) in terms of adequate gas sensing properties. NO$_x$ gases exhibited the most chemiresistive sensitivity towards the adsorbents. Other gases also showed significant chemiresistive sensitivity and distinct selectivity ratios, which would facilitate these alloys as chemiresistive sensors. Besides, noticeable work function variation ($\sim$20\%) of these systems manifested potential as work function based sensors. Cl$_2$ and NO$_2$ showed strong physical adsorption, while the rest of the gases were weakly to moderately physisorbed, resulting in very short recovery times (10$^{-1}$ $\sim$ 10$^{-6}$ seconds). Additionally, the distinctive absorption spectra observed for the gas analyte systems highlighted the potential of the proposed alloys as optical gas sensors. Temperature variation revealed that all gas molecules can be freed from the adsorbent BNGr-2 at 425 K. These findings imply hBN-Gr alloys as promising gas sensors for pollution auditing.} \\

\end{tabular}

 \end{@twocolumnfalse} \vspace{0.6cm}

  ]

\renewcommand*\rmdefault{bch}\normalfont\upshape
\rmfamily
\section*{}
\vspace{-1cm}


\footnotetext{\texorpdfstring{\textit{$^{a}$~Department of Electrical and Electronic Engineering, Bangladesh University of Engineering and Technology, Dhaka 1205, Bangladesh. E-mail: ahmedzubair@eee.buet.ac.bd}}}

\footnotetext{\dag~Electronic supplementary information (ESI) available. See DOI:}






\section{Introduction}
Fossil fuel combustion, industrial manufacturing, vehicle emissions, and many other sources are contributing to the increasing percentage of toxic gases in the atmosphere. To control this pollution, detection devices with high efficiency, sensitivity, and selectivity should be employed. Even a small concentration of some toxic gases such as Cl$_2$, CO, CO$_2$, NO, NO$_2$, and HCN can be unpleasant, poisonous, and harmful to the environment\,\cite{wang2018first,shahabi2017assessment}. Cl$_2$ is a pungent, yellowish-green, odorous gas that is frequently released from bleaching, sterilization, and water purification\,\cite{navale2017solid}. The occupational exposure limit (OEL) of Cl$_2$ gas is only 0.5 ppm, otherwise, it can cause skin irritation, suffocation, bronchospasm, and other severe symptoms\,\cite{navale2017solid}. Tasteless, colorless, and odorless CO gas is mostly produced by the incomplete combustion of fuels\,\cite{hong2019improved}.  It is one of the most dangerous gases, as human blood has a 250-300 times greater tendency to bond with CO than oxygen, resulting in the formation of carboxyhemoglobin (COHb). Therefore, affected individuals will experience unconsciousness, dizziness, and even death from severe exposure to CO\,\cite{samaranayake2024theoretical, naganaboina2021graphene}. Being a primary greenhouse gas, CO$_2$ is a threat to the environment, and it is also produced by combustion and industrial processes\,\cite{hunge2018multifunctional}. NO is mostly harmful to nature as it causes smog and acid rain. Besides, NO can irritate the respiratory system at high concentrations, and reacts with oxygen to form NO$_2$, which may lead to lung damage\,\cite{roohi2022adsorption}. NO$_2$ gas is a reddish-brown in color with a sharp and biting odor, which also causes respiratory inflammation and acid rain\,\cite{lee2018recent}. Another deadly gas is HCN, which can enter the human body through inhalation and also skin contact\,\cite{deji2021novel}. This colorless, faint bitter-almond-odorous gas molecule interferes with cellular respiration by blocking oxygen consumption, resulting in cytotoxic hypoxia, respiratory failure, and eventually death\,\cite{zhao2025first}. 
Graphene-like 2D materials have caught researchers’ attention in recent decades for various applications, including gas sensing, for their tunable electronic properties, high surface-to-volume ratio, high carrier mobility, and low power consumption\,\cite{abi2025functionalized,shahriar2022vacancy,bin2024first}. Silicene, germanene, stanene, transition metal dichalcogenides (TMDCs), III-V materials, metal oxides, and many other 2D materials have been investigated as hazardous gas sensors\,\cite{kumar2020room,sosa2022nh3, nguyen2022exploring,li2019advances}. Studies revealed that the introduction of dopants and defects significantly enhances the adsorption behavior of these materials\,\cite{baye2020graphene,baluchova2019porous}. \textit{Hou et al.} investigated cluster doping of Cr$_3$ into GaSe noticeably improved adsorption characteristics for Cl$_2$ and NO\,\cite{hou2022cr3}. \textit{Kim et al.} utilized the catalytic effect of Au by incorporating this metal with pristine WS$_2$ to improve interaction with CO gas molecules\,\cite{kim2020flexible}. Cr doping in TiO$_2$ was found to induce phase transformation, resulting in increased oxygen vacancies for better adsorption of CO$_2$. \textit{Linghu et al.} revealed that nonmetal doping, especially carbon doping, into MoS$_2$ breaks the bond of the NO$_2$ molecule when it comes into contact with C-doped 2H and 1T’-MoS$_2$, while pristine MoS$_2$ showed minimal interaction with NO$_2$\,\cite{linghu2019gas}. 

Hexagonal boron nitride (hBN) is renowned as ‘white graphene’ due to its structural similarity with graphene (Gr). hBN has been explored for various applications as a promising 2D material. Its high thermal conductivity, toxicological safety, and mechanical strength make it suitable for use in harsh environmental conditions, while its inherent lubricating characteristic broadens its potential applications\,\cite{int25,int26,int27}. Its ionic characteristics arise from the difference in electronegativity between B and N, resulting in strong interaction\,\cite{int28,int29}. In addition, its inert behavior makes it an ideal dielectric material to be used in cutting-edge electronic and optoelectronic devices\,\cite{int31}. The presence of weak van der Waals force between the adjacent layers of bulk hBN enables it to be peeled off. Although hBN has a wide band gap and poor electron transfer, which limits its electrochemical sensing performance, microstructural alteration enhances its sensing efficacy\,\cite{int29,int32}. Electrical and optical properties of hBN can be tailored via doping, substitution, functionalization, and hybridization. P-type (Be) or n-type (N,S, C, Si) doping can be introduced to control charge carriers and band gap tuning\,\cite{int24,int24A,int28,int30,jhumu2025hcn}. Carbon is the most advantageous among them as it has a similar lattice constant (1.45 {\AA} in hBN and 1.42 {\AA} in graphene, lattice mismatch is about 1.5\%) and structural layout\,\cite{int24,int30}. In a 2D graphene sheet, sp$^2$ hybridized carbon atoms are arranged in a honeycomb lattice\,\cite{int24A,int27}. The band structure of graphene shows a band gap of zero, as its valence band and conduction band meet at the Dirac point. A high amount of carrier mobility, thermal conductivity, tensile strength, and elastic modulus foster graphene to be broadly deployed in nanoelectronic devices. Its metallic behavior has the potential to be used to calibrate the electronic properties of pure hBN by using appropriate amounts of heteroatom doping of carbon. \textit{Ouyang et al.} characterized C-hBN by Fourier transform infrared (FTIR), photoluminescence (PL), and X-ray photoelectron spectroscopy (XPS), which created reactive sites. Consequently, the enhancement of electron transfer and electrocatalytic activity led to effective electrochemical determination of dopamine\,\cite{int29}. 
Band gap adaptation using vertical heterojunction of hBN and graphene can be achieved in the range of 0-114 meV, depending on the number of layers and configuration of layers\,\cite{torres2022band}. This limitation can be overcome using transverse heterojunctions. \textit{Zhang et al.} studied hBN-Gr transverse junctions and calculated band gaps of 0.86 eV to 1.177 eV for different percentages of hBN, which were utilized in Cl$_2$ gas sensing\,\cite{zhu2024electron}. Another study reveals that hBN/graphene lateral heterojunctions can be converted into a half-metallic or semiconducting material through interface engineering, e.g.,  modifying the interface configuration and graphene termination\,\cite{zhang2015interface}.  Moreover, carbon-enriched hBN crystals were investigated as an emerging van der Waals technology by \textit{Koperski et al.}\,\cite{koperski2020midgap}. It was used for enhancement of electronic, quantum photonics, and optoelectronic devices. \textit{Wang et al.} investigated that carbon doping increased the performance of hBN photodetectors up to 9.2 mA.W$^{-1}$ photoresistivity at 7.5\% carbon concentration, and performance deteriorates with further carbon percentages\,\cite{wang2020deep}. Carbon integration while growing hBN created unique shallow and deep levels and generated new optical transitions at distinct energies\,\cite{pelini2019shallow}. Moreover, carbon doping in boron nitride nanosheets (BNNs) facilitated Schottky barrier diode fabrication, which can be operated in high-temperature conditions\,\cite{int32}. C-doped hBN forming magnetic graphene islands performed as a successful catalyst for ORR (oxygen reduction reaction) and OER (oxygen evolution
reaction) through the investigation by \textit{Maji et al.}\,\cite{maji2019synergistic}. Behzad et al. revealed that double carbon doping, substituting boron and nitrogen atoms, resulted in more enhancement of electrical conductivity than single carbon doping in place of boron or nitrogen atoms\,\cite{int24A}. Another study showed that single doping of carbon into hBN enabled chemisorption of a few gases that were physisorbed when pristine hBN sheet was used as an adsorbent. However, adsorption behavior of gases such as H$_2$, N$_2$, CO, and NH$_3$ were physisorbed on both hBN and C-doped hBN\,\cite{int33}. High selectivity of NO$_x$ gases was achieved through employing in-plane graphene/hBN heterostructures as adsorbent material. Specific interfaces were found to be more potential for maximized adsorption, such as the CB interface for NO gas molecule, and the CN interface for NO$_2$\,\cite{de2019hybrid}. Although many studies have been conducted on the electronic, magnetic, optical, catalytic, and gas sensing properties of carbon-doped hBN, the effect of varying carbon concentration in hBN-graphene alloy and its gas sensing properties still remains largely unmapped.

This work inspects the consequences of carbon ratio variation in the electronic and gas sensitivity characteristics of hBN-graphene alloys with a uniform increment in the carbon ratio. The theoretical study has been carried out by employing density functional theory (DFT). First, electronic properties of eight types of alloys composed of hBN and graphene along with pure hBN and graphene monolayers, were investigated as prospective adsorbents. A wide range of adsorption sites were selected for these adsorbent structures to closely place the selected gas molecules, i.e., Cl$_2$, CO, CO$_2$, NO, NO$_2$, and HCN. Electronic parameters of the gas adsorbed systems were investigated through band structure, total density of states (TDOS), and partial density of states (PDOS) evaluation. The sensing properties were evaluated through conductivity, sensitivity, selectivity, work function, recovery time, and optical properties, which can foster future research on utilization of hBN-Gr alloy in environmental monitoring. In the end, temperature dependent adsorption energies of the most sensitive gas analyte systems were determined to examine the competency of the sensors to be used in variant temperature conditions.

\section{Computational Methods}

All calculations were performed using first-principles calculations based on density functional theory (DFT), implemented with CASTEP and Dmol3 modules of the Materials Studio 2020 software package. The generalized gradient approximation (GGA) was selected as the DFT exchange-correlation potential along with the Perdew-Burke-Ernzerhof (PBE) functional. Carbon atoms were introduced into a (3$\times$3$\times$1) supercell of an hBN monolayer. To have a thorough investigation of the impact of carbon proportion on the alloys,  2, 4, 6, 8, 10, 12, 14, and 16 carbon atoms were substituted into the B and N sites of the pristine hBN layer, referred as BNGr-1, BNGr-2, BNGr-3, BNGr-4, BNGr-5, BNGr-6, BNGr-7, and BNGr-8, which can be illustrated in Figs.\,\ref{fig:All_structures_opt}(b-i), respectively.  A 20 {\AA} vacuum slab was added along the z-direction to minimize interaction between periodic slabs. These structures were geometrically optimized until the convergence criteria of of maximum energy of $1\times10^{-6}$ Ha, maximum force of 0.002 Ha/{\AA}, and maximum displacement of 0.001{\AA} were satisfied. The DFT Semi-core Pseudopotential (DSPP) was employed to describe the core electrons, while a double numerical plus polarization (DNP) basis set was used for the valence electrons. Self-
consistent field (SCF) convergence was defined as $1\times10^{-6}$ along with 6$\times$6$\times$1 Monkhorst–Pack (k)-point mesh. 0.005 Ha thermal smearing was applied to the orbital occupation. For gas adsorption calculations, each gas molecule was initially positioned at different possible adsorption sites and orientations close to the surface of the adsorbents. Considered adsorption sites were top on atoms (C, B, and N), bridges ( C-C, B-C, B-N, and N-C), and hollows. The gas molecules were initially arranged in both parallel and perpendicular orientations with respect to the surface. The gas-surface systems were optimized, and the most energetically favorable systems were selected for subsequent calculations. Adsorption energy for the systems was calculated via the following expression, 
\begin{equation}
    E_{ad}=E_{gas/layer}-E_{gas}-E_{layer},
\end{equation}
where, E$_{ad}$, E$_{gas/layer}$, E$_{gas}$, and E$_{layer}$ are the adsorption energy, total energy of the gas-surface system, total energy of the isolated gas molecule, and total energy of the adsorbent layer, respectively. Hirshfeld charge transfer and electron density difference (EDD) were evaluated using the following expressions, 
\begin{equation}
\Delta Q = Q_{\mathrm{f}} - Q_{\mathrm{i}}
\label{eq:Hirshfeld}
\end{equation}
\begin{equation}\label{charge eqn}
\Delta\rho=\rho_{gas/layer}-\rho_{gas}-\rho_{layer}.
\end{equation}

In Eq.\,\ref{eq:Hirshfeld},  $\Delta Q$, Q$_{i}$, and Q$_{f}$ represent charge transfer, charge of the gas molecule before and after adsorption. In Eq.\,\ref{charge eqn}, $\Delta\rho$, $\rho_{gas/layer}$, $\rho_{gas}$, and $\rho_{layer}$ represent the electron density differences of the gas-adsorbed system, isolated gas molecule, and adsorbent layer, respectively. The adsorption distance was determined from the optimized geometry. Band structure, total density of states (TDOS), and partial density of states (PDOS) were calculated to investigate the electronic behavior of the systems before and after gas interaction. Work function, conductivity, sensitivity, selectivity, and recovery time were calculated afterwards. EDD and the optical absorption coefficient were calculated using the CASTEP module. Finally, adsorption energies of the most energetically preferable systems at varying temperature conditions were calculated. 

\section{Results and Discussion}
\subsection{Structural and electronic properties}

\begin{figure}[ht]
 \centering
 \includegraphics[width=0.5\textwidth]{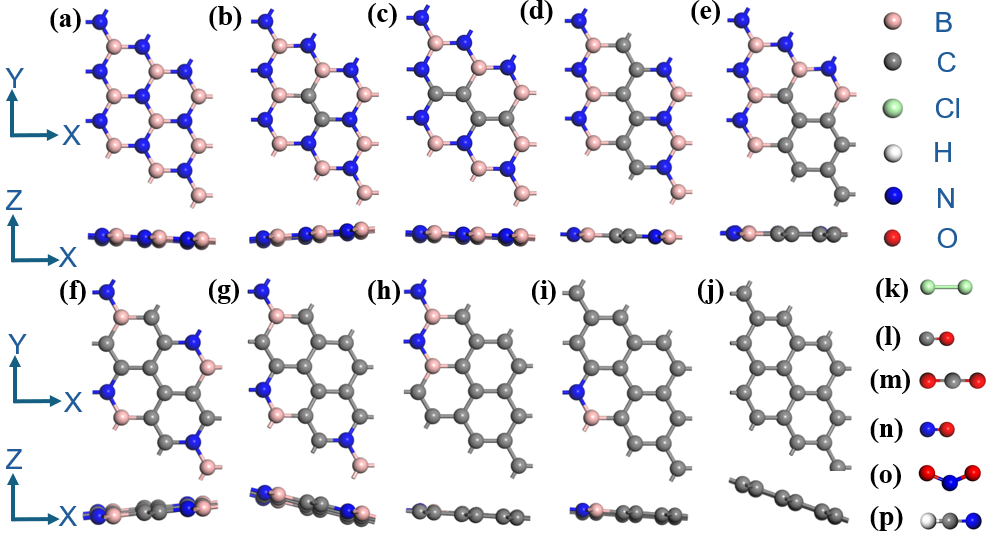}
 \caption{Geometry optimized (a) hBN, (b) BNGr-1 (11\% C), (c)  BNGr-2 (22\% C), (d) BNGr-3 (33\% C), (e) BNGr-4 (44\% C), (f) BNGr-5 (56\% C), (g) BNGr-6 (67\% C), (h) BNGr-7 (78\% C), (i) BNGr-8 (89\% C), and (j) Graphene, and gas molecules of (k) Cl$_2$, (l) CO, (m) CO$_2$, (n) NO, (o) NO$_2$, and (p) HCN.}
 \label{fig:All_structures_opt}
\end{figure}

\begin{figure*}
 \centering
 \includegraphics[width=\textwidth]{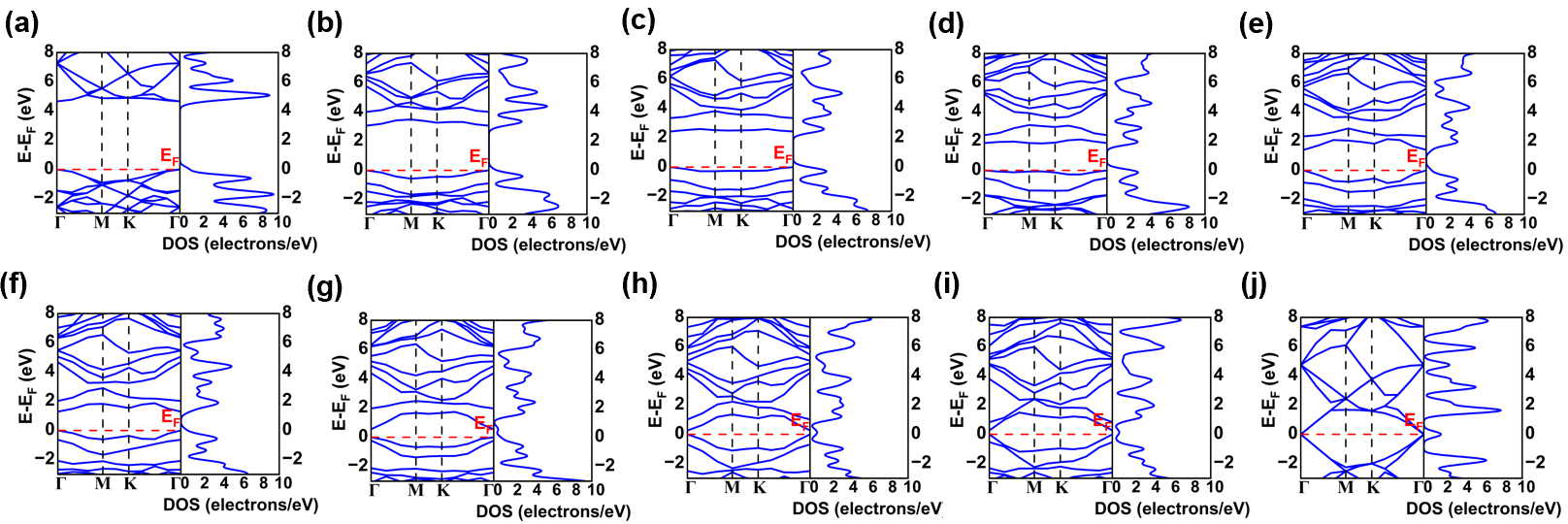}
 \caption{ Band structure and density of states (DOS) of (a) hBN monolayer, (b) BNGr-1, (c)  BNGr-2, (d) BNGr-3, (e) BNGr-4, (f) BNGr-5, (g) BNGr-6, (h) BNGr-7, (i) BNGr-8, and (j) Graphene.}
 \label{fig:Bandgap_and_DOS_of_all_hBN-Gr}
\end{figure*}

\begin{table*}[h]
\small
\caption{Lattice constant and band gap of pristine and modified hBN structures compared with graphene}
\label{tab:table_properties_of_hBN-Gr}
\begin{tabular*}{\textwidth}{@{\extracolsep{\fill}}lcccccccccc}
\hline
Property / Adsorbents & hBN & BNGr-1 & BNGr-2 & BNGr-3 & BNGr-4 & BNGr-5 & BNGr-6 & BNGr-7 & BNGr-8 & Graphene \\
\hline
Lattice constant (\AA) & 2.519 & 2.517 & 2.509 & 2.519 & 2.503 & 2.496 & 2.491 & 2.488 & 2.478 & 2.469 \\
Band gap (eV) & 4.645 & 3.057 & 2.436 & 1.830 & 1.350 & 1.318 & 0.246 & 0.294 & 0.283 & 0.000 \\
\hline
\end{tabular*}
\end{table*}

\begin{figure}[ht]
 \centering
 \includegraphics[width=0.5\textwidth]{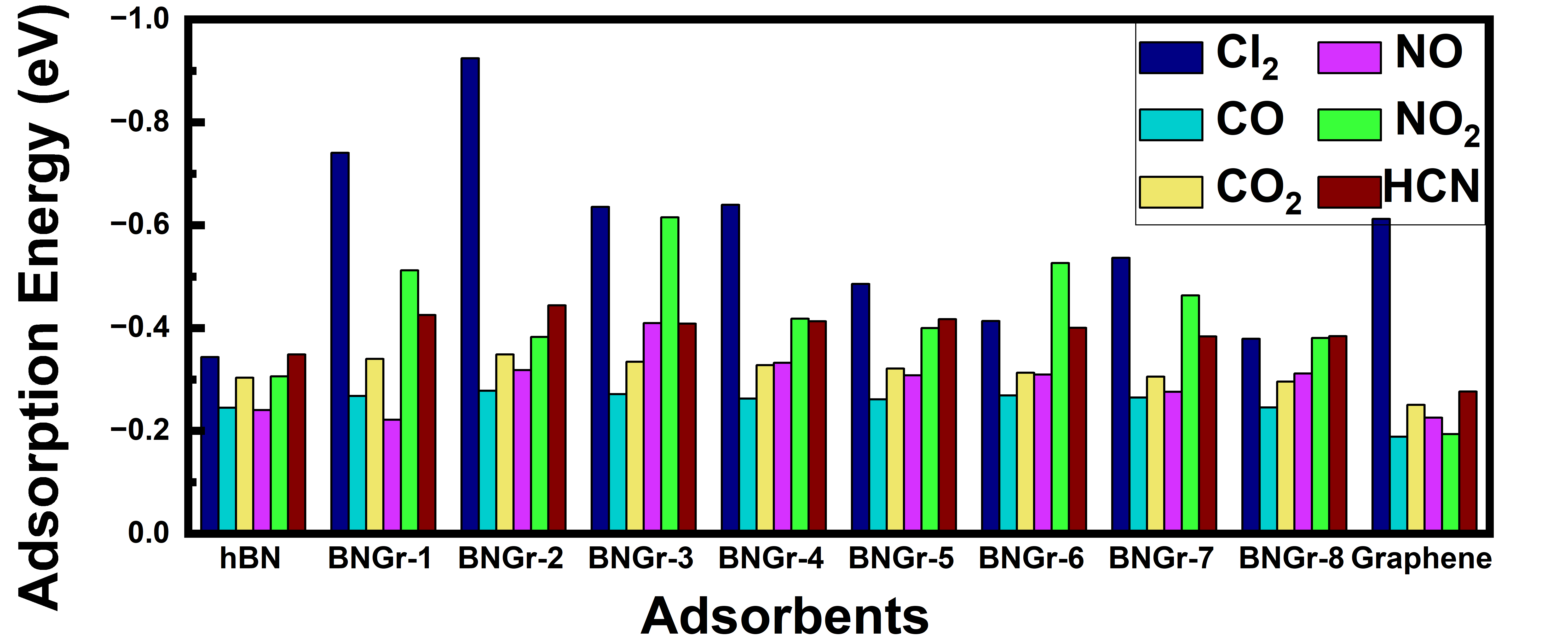}
 \caption{ Adsorption energy of gas molecules Cl$_2$, CO, CO$_2$, NO, NO$_2$, and HCN  on hBN-Gr alloys. }
 \label{fig:Adsorption_energy}
\end{figure}

Geometry optimized hBN and graphene monolayers had almost the same lattice parameters within honeycomb lattices, illustrated in Fig.\,\ref{fig:All_structures_opt}(a) and Fig.\,\ref{fig:All_structures_opt}(j) and presented in Table\,\ref{tab:table_properties_of_hBN-Gr}, which were consistent with previous studies\,\cite{renu2025interfacial,grison2026assessing}. These similarities predicted the potential and sustainability of developing alloys that can be synthesized from hBN and graphene. Optimized hBN-Gr alloys of variant carbon concentrations showed integrity with the pristine hBN and graphene sheets (Table\,\ref{tab:adsorption}). Significant changes occurred in the band structure and DOS of the alloys. An increased carbon ratio introduced impurity bands above the valence band and below the conduction band, and thus resulted in narrower band gaps. The higher the carbon ratio, the more the band gap was reduced, which can be observed from Fig.\,\ref{fig:Bandgap_and_DOS_of_all_hBN-Gr}. Corresponding DOS observations showed consistency with the band structures, as new peaks appeared between the band gap region. The contribution from carbon atoms transformed the wide band gap material hBN into a semiconductor, and at the highest concentrations it behaved like a metal.

\subsection{Adsorption configurations of gas molecules on hBN-Gr alloys}

\begin{figure}[ht]
 \centering
 \includegraphics[width=0.5\textwidth]{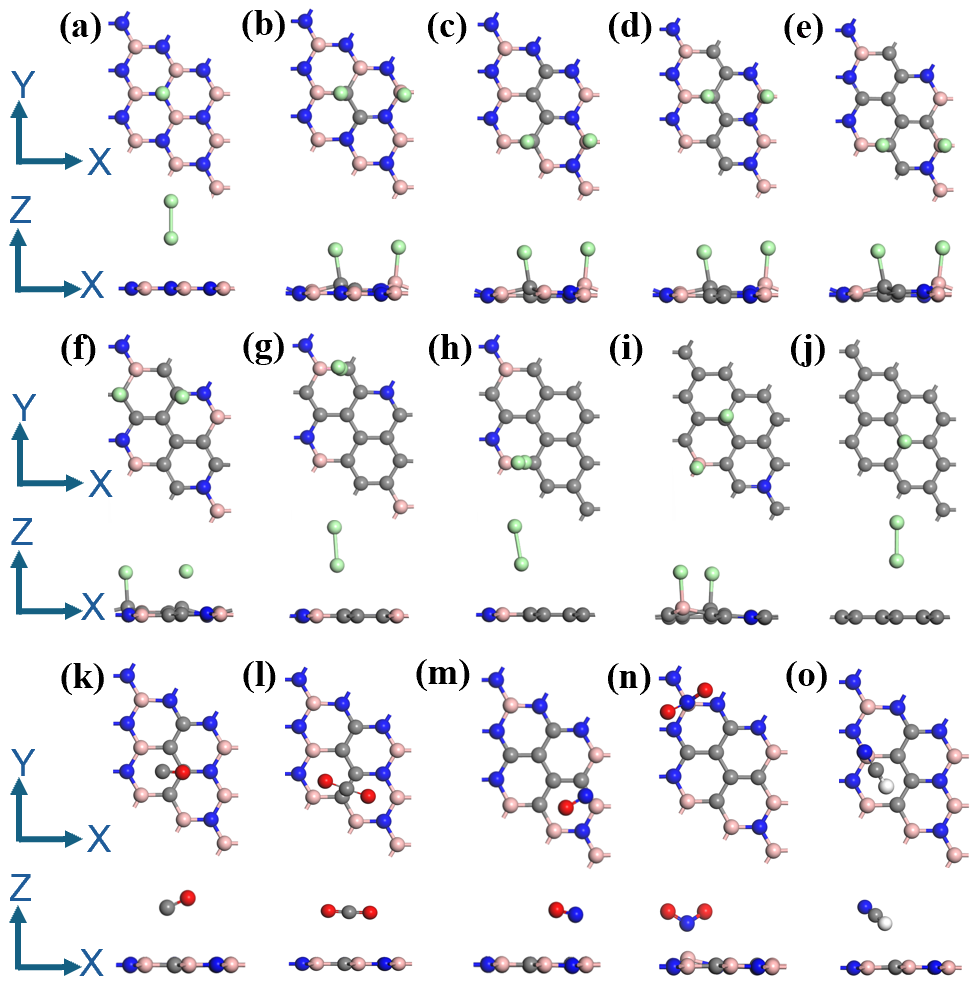}
 \caption{ Gas adsorbed systems of (a) Cl$_2$/hBN, (b) Cl$_2$/BNGr-1, (c) Cl$_2$/BNGr-2, (d) Cl$_2$/BNGr-3, (e) Cl$_2$/BNGr-4, (f) Cl$_2$/BNGr-5, (g) Cl$_2$/BNGr-6, (h) Cl$_2$/BNGr-7, (i) Cl$_2$/BNGr-8, (j) Cl$_2$/graphene, (k) CO/BNGr-2, (l) CO$_2$/BNGr-2, (m) NO/BNGr-3, (n) NO$_2$/BNGr-3, and (o) HCN/BNGr-2}
 \label{fig:gas_ads_systems}
\end{figure}

Geometry-optimized gas molecules were placed near the adsorbents, and the interaction strength was evaluated by calculating adsorption energies. Adsorbed systems were optimized, and the most energetically favorable systems were considered for further calculations. The most stable adsorbed sites for Cl$_2$/alloys and the highest adsorbed systems for other gases such as CO/BNGr-2, CO$_2$/BNGr-2, NO/BNGr-3, NO$_2$/BNGr-3, and HCN/BNGr-2 systems can be observed from Fig.\,\ref{fig:gas_ads_systems}. Adsorption characteristics such as adsorption energy (E$_{ad}$), adsorption distance (D), Hirshfeld charge transfer ($\Delta$Q), band gap before (E$_{gi}$) and after (E$_{gf}$) gas adsorption are listed in Table\,\ref{tab:adsorption}. The lowest adsorption energy (maximum in absolute value) indicates the strongest interaction with the adsorbents. Hence, BNGr-2 and BNGr-3 were revealed to be the best adsorbents for the observed gas molecules. However, all the calculated adsorption processes were exothermic, as indicated by their negative adsorption energies, suggesting stability. For chemisorption, the required adsorption energy is $-$1 eV or less\,\cite{bin2024first}. The most favorable adsorption for each gas has been presented in Fig.\,\ref{fig:gas_ads_systems}. As seen in Fig.\,\ref{fig:gas_ads_systems} for the best gas adsorptions, gas molecules were adsorbed in various orientations, and the best adsorption sites were not fixed at all; gas molecules favored different adsorption sites. Adsorption traits of the selected hBN-Gr alloys for each gas molecule are discussed in the following sections. 

\begin{table*}[h]
\fontsize{7pt}{12pt}\selectfont
\setlength{\tabcolsep}{1pt}
\caption{\ Adsorption energy ($E_{\mathrm{ad}}$), adsorption distance ($D$), and charge transfer ($\Delta Q$) of gas molecules on hBN-based adsorbents}
\label{tab:adsorption}
\resizebox{\textwidth}{!}{%
\begin{tabular*}{\textwidth}{@{\extracolsep{\fill}}lcccccccccccccccccc}
\hline
\multirow{3}{*}{\centering{Adsorbent}} &
\multicolumn{18}{c}{{Adsorbed Gas Molecules}} \\
\cmidrule(lr){2-19}
& \multicolumn{3}{c}{{Cl$_2$}}
& \multicolumn{3}{c}{{CO}}
& \multicolumn{3}{c}{{CO$_2$}}
& \multicolumn{3}{c}{{NO}}
& \multicolumn{3}{c}{{NO$_2$}}
& \multicolumn{3}{c}{{HCN}} \\
\cmidrule(lr){2-4} \cmidrule(lr){5-7} \cmidrule(lr){8-10}
\cmidrule(lr){11-13} \cmidrule(lr){14-16} \cmidrule(lr){17-19}
& {$E_{\mathrm{ad}}$ (eV)} & {$D$ (\AA)} & {$\Delta Q$ ($|e|$)}
& {$E_{\mathrm{ad}}$ (eV)} & {$D$ (\AA)} & {$\Delta Q$ ($|e|$)}
& {$E_{\mathrm{ad}}$ (eV)} & {$D$ (\AA)} & {$\Delta Q$ ($|e|$)}
& {$E_{\mathrm{ad}}$ (eV)} & {$D$ (\AA)} & {$\Delta Q$ ($|e|$)}
& {$E_{\mathrm{ad}}$ (eV)} & {$D$ (\AA)} & {$\Delta Q$ ($|e|$)}
& {$E_{\mathrm{ad}}$ (eV)} & {$D$ (\AA)} & {$\Delta Q$ ($|e|$)} \\
\midrule
hBN      & $-$0.344 & 2.758 &  0.131 & $-$0.245 & 3.093 & $-$0.003 & $-$0.303 & 3.017 &  0.009 & $-$0.241 & 2.933 &  0.001 & $-$0.306 & 2.896 & $-$0.022 & $-$0.348 & 2.783 & $-$0.018 \\
BNGr-1     & $-$0.741 & 1.876 & $-$0.231 & $-$0.268 & 3.015 & $-$0.017 & $-$0.340 & 2.986 & $-$0.018 & $-$0.222 & 1.780 & $-$0.036 & $-$0.512 & 2.570 & $-$0.136 & $-$0.425 & 2.503 & $-$0.049 \\
BNGr-2     & $-$0.925 & 1.885 & $-$0.231 & $-$0.278 & 2.873 & $-$0.028 & $-$0.349 & 3.302 & $-$0.021 & $-$0.318 & 2.697 & $-$0.057 & $-$0.383 & 3.194 & $-$0.163 & $-$0.444 & 2.366 & $-$0.065 \\
BNGr-3     & $-$0.635 & 1.890 & $-$0.219 & $-$0.271 & 2.887 & $-$0.025 & $-$0.334 & 2.887 & $-$0.016 & $-$0.409 & 2.466 & $-$0.038 & $-$0.615 & 1.726 & $-$0.212 & $-$0.409 & 2.754 & $-$0.030 \\
BNGr-4     & $-$0.640 & 1.914 & $-$0.118 & $-$0.262 & 2.995 & $-$0.017 & $-$0.328 & 3.023 & $-$0.020 & $-$0.332 & 2.560 & $-$0.059 & $-$0.418 & 1.852 & $-$0.201 & $-$0.413 & 2.535 & $-$0.045 \\
BNGr-5     & $-$0.486 & 1.915 & $-$0.232 & $-$0.261 & 2.998 & $-$0.028 & $-$0.321 & 3.027 & $-$0.015 & $-$0.308 & 2.656 & $-$0.035 & $-$0.400 & 2.219 & $-$0.171 & $-$0.417 & 2.855 & $-$0.025 \\
BNGr-6     & $-$0.413 & 2.746 & $-$0.115 & $-$0.269 & 3.142 & $-$0.019 & $-$0.313 & 3.051 & $-$0.018 & $-$0.310 & 2.623 & $-$0.019 & $-$0.526 & 1.734 & $-$0.214 & $-$0.400 & 2.552 & $-$0.041 \\
BNGr-7     & $-$0.537 & 2.736 & $-$0.089 & $-$0.265 & 3.138 & $-$0.015 & $-$0.305 & 3.073 & $-$0.015 & $-$0.276 & 2.730 & $-$0.010 & $-$0.463 & 1.519 & $-$0.185 & $-$0.383 & 2.742 & $-$0.043 \\
BNGr-8     & $-$0.452 & 1.950 & $-$0.196 & $-$0.246 & 2.937 & $-$0.014 & $-$0.296 & 3.056 & $-$0.018 & $-$0.312 & 2.642 & $-$0.007 & $-$0.381 & 1.867 & $-$0.180 & $-$0.384 & 2.937 & $-$0.014 \\
Graphene & $-$0.612 & 3.013 & $-$0.056 & $-$0.189 & 3.086 & $-$0.019 & $-$0.250 & 3.055 & $-$0.022 & $-$0.226 & 2.882 & $-$0.006 & $-$0.194 & 2.896 & $-$0.140 & $-$0.276 & 2.857 & $-$0.033 \\
    \hline
\end{tabular*}}
\end{table*}

\subsubsection{Adsorption behavior of \texorpdfstring{Cl$_2$}{Cl2}}

\begin{figure}[ht]
 \centering
 \includegraphics[width=0.5\textwidth]{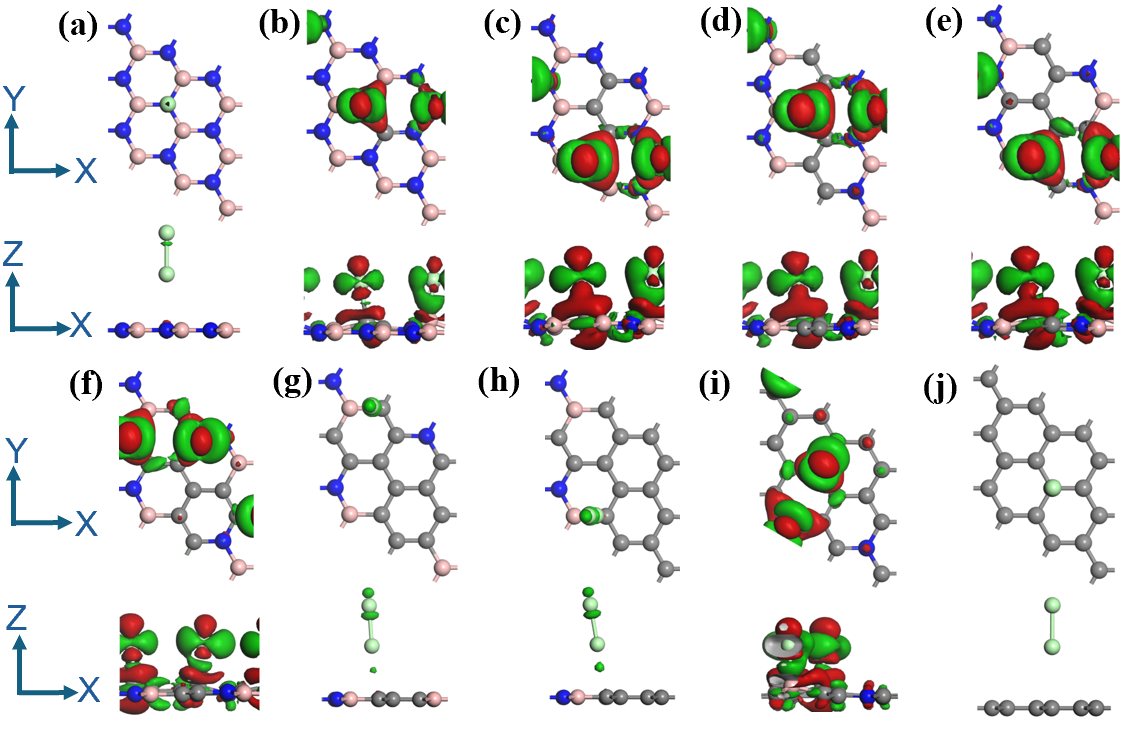}
 \caption{ Electron Density Difference of Cl$_2$ adsorbed (a) hBN, (b) BNGr-1, (c) BNGr-2, (d) BNGr-3, (e) BNGr-4, (f) BNGr-5, (g) BNGr-6, (h) BNGr-7, (i) BNGr-8, and (j) Graphene. Here the green (red) color denotes electron accumulation (depletion). Isovalue is 0.025 electrons per $\text{\AA}^{-3}$.}
 \label{fig:EDD_Cl2_hBNGr}
\end{figure}

Cl$_2$ showed the strongest interaction with the adsorption sheets among the gas molecules investigated. The highest adsorption energy was calculated as $-$0.93 eV when the Cl$_2$ gas molecule interacted with BNGr-2. The bond between Cl-Cl atoms of the horizontally placed gas molecule broke down, and the Cl atoms were found to create bonds with carbon and boron atoms of the adsorbent sheet, bending the sheet structure towards the gas atoms. Eventually, the adsorption distance became very small as 1.89 {\AA}. Charge transfer of $-$0.23 e indicated strong adsorption, which can be observed from Fig.\,\ref{fig:EDD_Cl2_hBNGr}(c). The general adsorption behavior of Cl$_2$ gas with the adsorbent sheets unfolded that the interaction increased with increasing carbon ratio in the alloys and reached a maximum point for the carbon percentage of 22\% (Table\,\ref{tab:adsorption}). A decrease in adsorption energy and charge transfer, and an increase in adsorption distance became evident for higher carbon ratios. However, some fluctuations were observed for BNGr-7, BNGr-8, and graphene. 

\subsubsection{Adsorption behavior of CO}
The carbon atom of the CO molecule showed an inclination towards the adsorbent sheets, whether it was placed horizontally or vertically. The adsorption site bridge B-C was found to be the most favorable for adsorption for the highest number of times, although maximum adsorption occurred at the site top on B of BNGr-2. CO interacted the most with BNGr-2, with an adsorption energy of $-$0.28 eV, though adsorption on other alloys was not significantly different. However, there was a similar trend to that for other gas adsorptions that increasing carbon ratio improved adsorption behavior until 22\%, then the adsorption energies gradually decreased (Table\,\ref{tab:adsorption}). The adsorption distance also was the lowest for adsorption on BNGr-2, which was 2.87 \AA. Hirshfeld charge transfer value supported the maximum adsorption on BNGr-2, as the value was $-$0.028 e, the highest for CO adsorption. A net result of charge accumulation can be observed from the EDD calculations of CO gas-adsorbed systems (provided in the ESI). The plots validated the negative value of charge transfer around the CO gas molecule. Negligible charge transfer can be observed for the alloys that had higher ratios of carbon. Nonetheless, physisorption was indicated by the low adsorption energy and charge transfer.  

\subsubsection{Adsorption behavior of \texorpdfstring{CO$_2$}{CO2}}

From Fig.\,\ref{fig:Adsorption_energy}, it can be observed that maximum adsorption for CO$_2$ was achieved for interaction with BNGr-2 ($-$0.35 eV). The most stable site for this case was the bridge C-C. Initially, the CO$_2$ molecule was placed vertically, but it was found to rotate to a horizontal position due to adsorption, with an adsorption distance of 3.30 {\AA} and a charge transfer of $-$0.021 e. These findings can be illustrated from the electron density difference mapping of CO$_2$/alloys (provided in the ESI). Negative charge depletion from the sheet supported the negative charge transfer value, and the minimal amount of charge transfer indicated physisorption. Other alloys exhibited nearly the same amount of adsorption properties while maintaining the trait of increased adsorption performance with increased carbon ratio, attaining the highest interaction for BNGr-2, and then performance declined with higher carbon ratios, which was analogous to the behavior of the previous gases. The Hirshfeld charge transfer calculated for the CO$_2$ molecule was negative for all systems except pristine hBN. A very low amount of positive charge transfer occurred as a result of very low charge discharge from hBN, indicating poor interaction. Charge depletion from the gas molecules occurred in other systems due to the adequate amount of charge loss from the carbon atom of CO$_2$. This charge loss was observed to decrease with the growing carbon proportion in the alloys and ultimately reduced interaction. 

\subsubsection{Adsorption behavior of NO}

Adsorption energies of NO/hBN-Gr increased with increase in carbon concentration, and at a 33\% carbon ratio it reached a peak of $-$0.41 eV. Adsorption energy decreased thereafter, and NO/graphene exhibited the weakest interaction with an adsorption energy of $-$0.23 eV. Hirshfeld charge transfer and the adsorption distance fluctuated for different carbon ratios. Positive charge accumulated on the adsorbent surfaces (See the ESI for the detailed formulation), which corroborated the negative charge transfer value of the NO molecule. Charge accumulation was very low for the alloys with very high carbon percentages. NO/BNGr-3 showed a charge transfer of $-$0.038 e and a very low adsorption distance of 1.91 {\AA}, which was reflected in the EDD calculation (provided in the ESI). However, physisorption was identified for NO adsorption. 

\subsubsection{Adsorption behavior of \texorpdfstring{NO$_2$}{NO2}}

\begin{figure}[ht]
 \centering
 \includegraphics[width=0.5\textwidth]{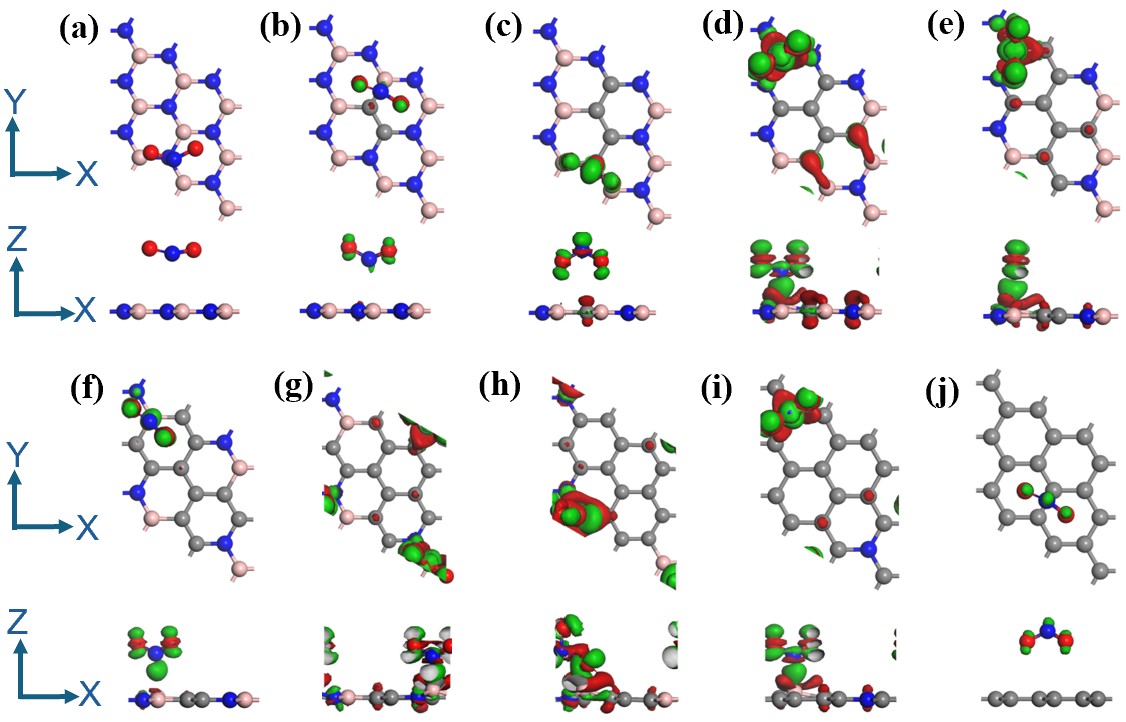}
 \caption{ Electron Density Difference of NO$_2$ adsorbed (a) hBN monolayer, (b) BNGr-1, (c) BNGr-2, (d) BNGr-3, (e) BNGr-4, (f) BNGr-5, (g) BNGr-6, (h) BNGr-7, (i) BNGr-8, and (j) Graphene. Here, the green (red) color denotes electron accumulation (depletion). Isovalue is 0.025 electrons per $\text{\AA}^{-3}$.}
 \label{fig:EDD_NO2_hBNGr}
\end{figure}

Adsorption features of NO$_2$ were similar to those of the NO molecule, and it also interacted the most with BNGr-3. For this system, the adsorption energy is $-$0.62 eV, which was noticeably higher than that of NO/BNGr-3. An adsorption distance of 1.73 {\AA} was significantly low, which promoted stronger adsorption. Charge transfer of $-$0.21 e and EDD from Fig.\,\ref{fig:EDD_NO2_hBNGr}(d) shows a sufficient depletion of charge from the NO$_2$ molecule and accumulation in the adsorbents. All of these results suggested strong physical adsorption of the NO$_2$ molecule in the BNGr-3 adsorbent. Smaller adsorption energies, larger adsorption distances, and smaller charge transfers were observed for other alloy-based systems (Table\,\ref{tab:adsorption}). For NO$_2$/hBN, the adsorption energy was calculated $-$0.31 eV, the adsorption distance was 2.89 {\AA}, and the charge transfer was $-$0.022 e, which indicated a moderate adsorption with weak adsorption properties. Subsequently, gradually improved adsorption behavior towards NO$_2$ was observed with increasing carbon fraction up to 33\%. Afterward, adsorption energy dropped steadily, and the NO$_2$/Graphene system showed a low adsorption energy of $-$0.19 eV along with a moderate charge transfer of $-$0.14 e. These responses can be illustrated from Fig.\,\ref{fig:EDD_NO2_hBNGr} through EDD of NO$_2$/hBNGr systems. The green area, which was denoted as electron accumulation, increased from the adsorbents BNGr-2 to BNGr-5 and slowly diminished for higher carbon ratios. 

\subsubsection{Adsorption behavior of HCN}

HCN molecules were physiosorbed on the hBN-Gr alloys, as the adsorption energies were not sufficient for chemisorption. However, BNGr-2 appeared to be the most energetically stable system for adsorption. Adsorption energy, distance, and Hirshfeld charge transfer were calculated as $-$0.44 eV, 2.37 {\AA}, and $-$0.065 e. EDD analysis of the system (provided in the ESI) shows electron depletion from the BNGr-2 surface and accumulation near hydrogen and carbon atoms of the HCN molecule. Other systems showed similarity with Cl$_2$, CO, and CO$_2$ gas adsorption traits, as adsorption energy reached a maximum point at 22\% carbon percentage, and decreased afterwards. The least favorable adsorption energy value was calculated for HCN/Graphene as $-$0.28 eV along with a negligible charge transfer of $-$0.033 e. The high local electron concentration near the HCN molecules aligned well with the negative charge transfer values for HCN/hBNGr in Table \,\ref{tab:adsorption}. However, the HCN molecule was observed to tilt slightly, and the H atom became the closest atom after adsorption, regardless of whether the molecule was placed horizontally or vertically.

\subsection{ Electronic structure of gas-adsorbed systems}

\begin{figure}[ht]
 \centering
 \includegraphics[width=0.5\textwidth]{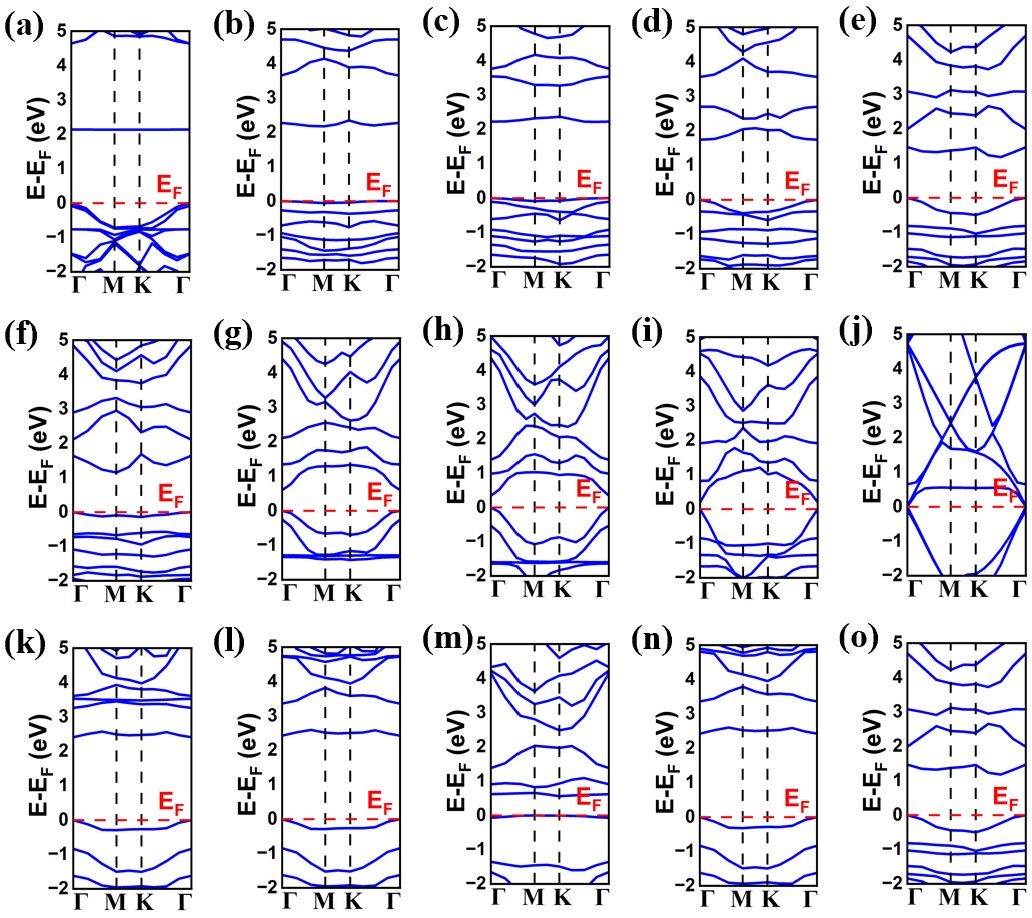}
 \caption{ Band structure of gas adsorbed systems of (a) Cl$_2$/hBN, (b) Cl$_2$/BNGr-1, (c) Cl$_2$/BNGr-2, (d) Cl$_2$/BNGr-3, (e) Cl$_2$/BNGr-4, (f) Cl$_2$/BNGr-5, (g) Cl$_2$/BNGr-6, (h) Cl$_2$/BNGr-7, (i) Cl$_2$/BNGr-8, (j) Cl$_2$/graphene, (k) CO/BNGr-2, (l) CO$_2$/BNGr-2, (m) NO/BNGr-3, (n) NO$_2$/BNGr-3, and (o) HCN/BNGr-2.}
 \label{fig:Bandgap_Cl2_hBNGr}
\end{figure}

To further study the extent of interaction of gas molecules with the adsorbent alloys, the band structure, TDOS, and PDOS of the gas-adsorbed systems were analyzed. The proximity of the gas molecules was expected to influence the electronic properties of the pristine structures. Figs.\,\ref{fig:Bandgap_Cl2_hBNGr}(a)-(j) show the band structures of Cl$_2$-adsorbed structures. It was evident that the presence of the Cl$_2$ molecule profoundly altered the band structures of the pristine ones. The band gap reduced noticeably, resulting in the appearance of bands under the conduction band. Flat lines appeared near the Fermi level in the presence of Cl$_2$ for alloys with the smaller carbon ratios (Fig.\,\ref{fig:Bandgap_Cl2_hBNGr}(a)-(c)). For the alloys with higher-order carbon ratios, the bands were distorted in shape. However, band gaps fluctuated minimally for these systems. As a result, the TDOS of these structures distorted, showing a clear reduction of the band gap for the gas-adsorbed systems. Slight overlap between the p orbitals of Cl and N atoms was visible for all systems in the $-$2 to $-$0.5 eV range. However, strong hybridization between Cl and adsorbent atoms C and B was evident. The most favorable system, Cl$_2$/BNGr-3, had pronounced overlap between the p orbital of B, p orbital of C, and p orbital of Cl in the range of 2 to 4 eV. This indicated strong bonding and bond formation with the Cl atoms in  Fig.\,\ref{fig:gas_ads_systems}(c), which also aligned with this observation.

Band structures of CO/alloys are provided in the ESI. The influence of CO on the adsorbents of lower carbon ratios moderately reduced the band gaps, and manifested weak adsorption. However, the strongest interaction among these systems, which was CO/BNGr-2, had a band gap of 2.409 eV (0.54\% reduction from the band gap of pristine BNGr-2, which was 2.436 eV). The band gap increased marginally for the higher carbon ratio structures, which was apparent in the TDOS plots provided in the ESI. Orbitals of the C and O atoms of the CO molecule showed affinity for p orbitals of B and N, and overlap occurred between 2.5 eV and 4 eV. The PDOS of the gas molecule atoms was almost stagnant, while B-p and N-p were distorted and slightly shifted towards C and O of CO.

Band structure of the adsorbent materials was not affected significantly in case of CO$_2$ adsorption. The introduction of CO$_2$ gas molecules did not instigate new states yet distorted the band shapes. Hence, the pristine structures dominated the band structures of the gas-adsorbed systems. The band gap was reduced or expanded by a very small amount, which indicated weak influence on the adsorbents. It was evident from the DOS calculations of CO$_2$/hBN-Gr that the p orbitals of C and O atoms hybridized with the p orbitals of C and B atoms of the adsorbents in the range of 4$-$5 eV. No significant differences were observed, although the adsorbents were different alloys.

\begin{figure*}
 \centering
 \includegraphics[width=\textwidth]{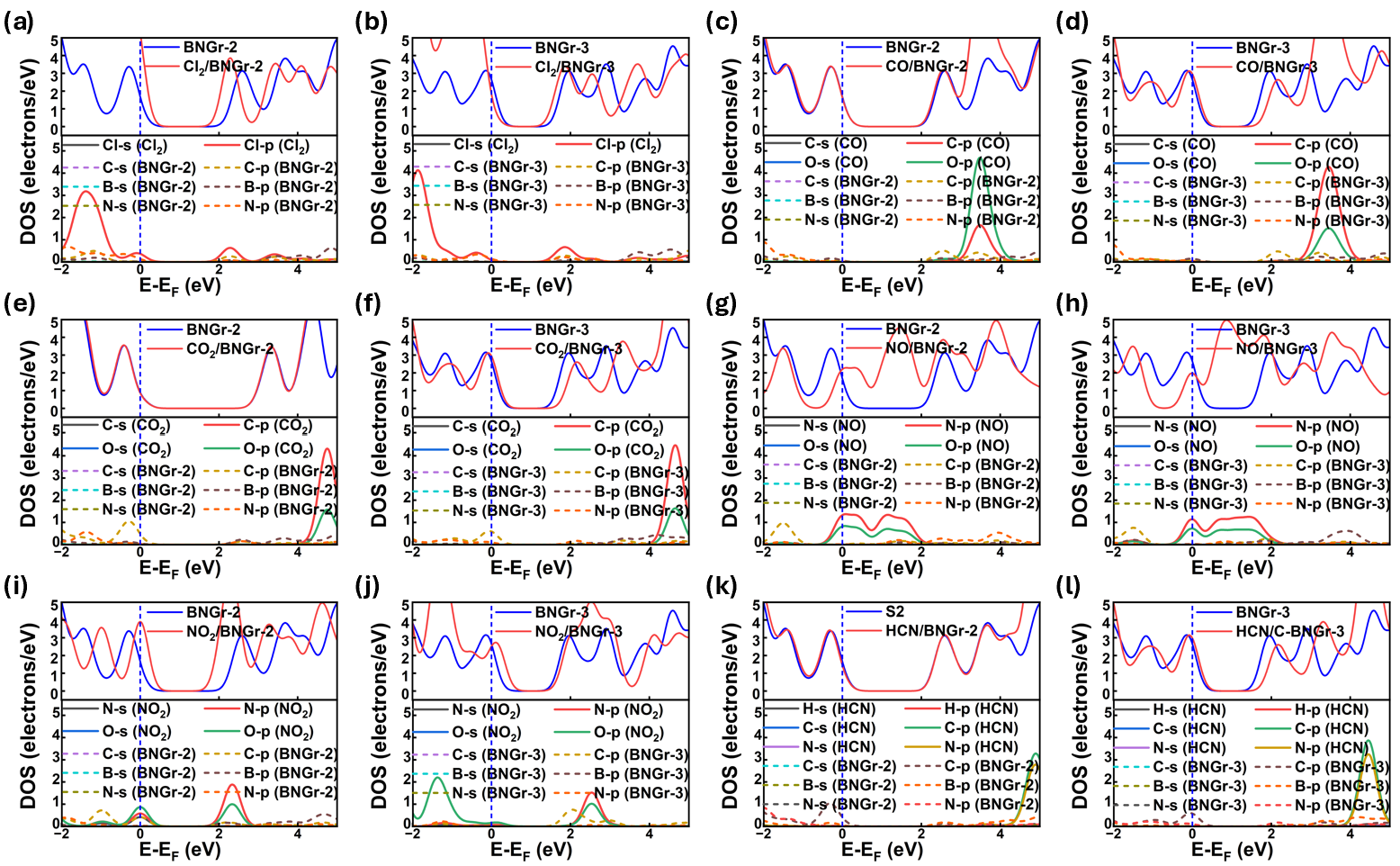}
 \caption{ DOS and PDOS of (a) Cl$_2$/BNGr-2, (b) CO/BNGr-2, (c) CO$_2$/BNGr-2, (d) NO/BNGr-2, (e) NO$_2$/BNGr-2, (f) HCN/BNGr-2, (g) Cl$_2$/BNGr-3, (h) CO/BNGr-3, (i) CO$_2$/BNGr-3, (j) NO/BNGr-3, (k) NO$_2$/BNGr-3, and (l) HCN/BNGr-2.}
 \label{fig:all_DOS_PDOS}
\end{figure*}

Band gap modification was prominent for NO gas adsorption on the alloys. Band gap values almost dropped to zero for these systems. The arrival of new states due to NO interaction demonstrated a pronounced influence of NO on the electronic characteristics. About 87\% and 70\% reduction occurred in the band gap of BNGr-2 and BNGr-3, respectively, which would induce high sensitivity towards NO gas molecules. The effect of NO gas exposure on the alloys was very distinct from the TDOS of NO/alloys. TDOS and PDOS calculations of NO/BNGr-2, NO/BNGr-3, and NO/BNGr-4 revealed that the states of both the gas molecule and the adsorbents were modified and affected each other. The p orbitals of N and O atoms from NO and the p orbital of B from BNGr-3 overlapped in the 1$-$2 eV region, indicating engagement with the gas molecule(Fig.\,\ref{fig:all_DOS_PDOS} (h)). PDOS of other gas-adsorbed systems exhibited weak interaction, except NO/BNGr-8, which exhibited a bit stronger interaction through moderate hybridization between N-p (from NO), O-p (from NO), and N-p (from BNGr-8) in the 0.5$-$2 eV region (details provided in the ESI). 

The adsorption of NO$_2$ noticeably altered the band structure of the pristine adsorbents. The band structures of NO$_2$/alloy systems were heavily dominated by the gas molecule. About 1.85 eV of band energy was reduced for NO$_2$/BNGr-3 as the conduction band minimum appeared very close to the Fermi level (see ESI for details). A huge band gap reduction was apparent in the other NO$_2$ gas-adsorbed systems, also. NO$_2$ adsorption gave rise to new states around 2$-$3 eV which contributed to narrowing the band gap. Consequently, the TDOS of NO$_2$/alloys were strongly influenced by the gas molecule. Strong hybridization was observed for the p orbitals of N and O atoms with p orbitals of the C atom of the alloy BNGr-3, which yielded strong physisorption of NO$_2$. A moderate amount of hybridization of the p orbitals of N and O atoms and the B atom of BNGr-1 and BNGr-2 was noticed. In the case of NO$_2$ interaction with higher carbon ratio alloys (BNGr-4, BNGr-5, BNGr-6, BNGr-7, and BNGr-8), N and O from the gas molecule marginally hybridized with the carbon of the alloys. 

HCN adsorption marginally affected the band structures of the pristine alloys. However, the band gap was increased after gas adsorption for some alloys, which was rare for the other gases. Widening of the band gap was detected for BNGr-1, BNGr-3, BNGr-4, BNGr-5, and BNGr-6 due to the influence of the HCN gas molecule. The slight variations in the band structures of HCN/alloys were perceivable from the TDOS calculations (see ESI for details). The adsorbed HCN molecule did not incorporate new states. Hence, characteristics of the pristine structures were prevalent in the TDOS of the gas-adsorbed structures. PDOS calculations of HCN/alloys showed that the B atom of the alloys interacted with the C and N atoms of the HCN molecule moderately. In the range of 4$-$6 eV, boron p-orbitals hybridized with the nitrogen p- and carbon p-orbitals of HCN when BNGr-2 and BNGr-3 were taken as the adsorbent (see Figs.\,\ref{fig:all_DOS_PDOS} (k)-(l)).

\subsection{Gas Sensing Performance of hBN-Gr Alloys}

\begin{table*}
\fontsize{9pt}{12pt}\selectfont
\setlength{\tabcolsep}{1pt}
  \caption{\ Work function before ($\Phi_i$) and after ($\Phi_f$) gas adsorption, and  Work function variation ($\Delta\Phi$) of hBN–Gr alloys after gas adsorption}
\label{tab:work_function}

\begin{tabular*}{\textwidth}{@{\extracolsep{\fill}}l c cc cc cc cc cc cc}
\hline
\multirow{2}{*}{Adsorbent} &
\multirow{2}{*}{$\Phi_i$ (eV)} &
\multicolumn{2}{c}{Cl$_2$} &
\multicolumn{2}{c}{CO} &
\multicolumn{2}{c}{CO$_2$} &
\multicolumn{2}{c}{NO} &
\multicolumn{2}{c}{NO$_2$} &
\multicolumn{2}{c}{HCN} \\
\cmidrule(lr){3-4}\cmidrule(lr){5-6}\cmidrule(lr){7-8}
\cmidrule(lr){9-10}\cmidrule(lr){11-12}\cmidrule(lr){13-14}
 &  &
 $\Phi_f$ (eV) & $\Delta\Phi$ (\%) &
 $\Phi_f$ (eV) & $\Delta\Phi$ (\%) &
 $\Phi_f$ (eV) & $\Delta\Phi$ (\%) &
 $\Phi_f$ (eV) & $\Delta\Phi$ (\%) &
 $\Phi_f$ (eV) & $\Delta\Phi$ (\%) &
 $\Phi_f$ (eV) & $\Delta\Phi$ (\%) \\
\hline
hBN      & 5.959 & 6.150 & 3.20 & 5.959 & 0.00 & 5.932 & $-$0.46 & 4.463 & $-$25.11 & 5.878 & $-$1.37 & 5.932 & 0.46 \\
BNGr-1       & 5.143 & 6.150 & 19.58 & 5.388 & 4.76 & 5.415 & 5.29 & 5.143 & 0.00 & 5.415 & 5.29 & 5.524 & 7.41 \\
BNGr-2       & 4.925 & 6.123 & 24.31 & 5.252 & 6.63 & 5.225 & 6.08 & 5.170 & 4.97 & 5.034 & 2.21 & 5.442 & 10.50 \\
BNGr-3       & 4.789 & 5.905
         & 23.30
         & 5.143 & 7.39 & 5.116 & 6.82 & 3.864 & $-$19.32 & 5.497 & 14.77 & 5.034 & 5.11 \\
BNGr-4       & 4.381 & 5.606 & 27.95 & 4.980 & 13.66 & 4.762 & 8.70 & 4.027 & $-$8.07 & 5.034 & 14.91 & 5.061 & 15.53 \\
BNGr-5       & 4.789 & 5.769 & 20.45 & 5.089 & 6.25 & 4.789 & 0.00 & 4.354 & $-$9.09 & 5.279 & 10.23 & 4.898 & 2.27 \\
BNGr-6       & 4.435 & 5.442 & 22.70 & 5.034 & 13.50 & 4.980 & 12.27 & 4.490 & 1.23 & 5.606 & 26.38 & 5.061 & 14.11 \\
BNGr-7       & 4.708 & 5.361 & 13.87 & 4.898 & 4.05 & 4.871 & 3.47 & 4.327 & $-$8.09 & 5.524 & 17.34 & 4.708 & 0.00 \\
BNGr-8       & 4.708 & 5.442 & 15.61 & 4.272 & $-$9.25 & 4.816 & 2.31 & 4.272 & $-$9.25 & 5.170 & 9.83 & 4.898 & 4.05 \\
Graphene & 4.789 & 5.170 & 7.95 & 5.007 & 4.55 & 4.980
          & 3.98
          & 4.327 & $-$9.66 & 5.034 & 5.11 & 4.327 & $-$9.66 \\
\hline
\end{tabular*}
\end{table*}

\subsubsection{Work Function Variation}
The variation of the work function depicts the surface sensitivity of a sensor. The work function ($\phi$) can be determined by the energy difference between the electrostatic potential in vacuum ($V_{\infty}$) and the energy at the Fermi level ($E_\mathrm{F}$)\,\cite{lin2023dft}, which is expressed by the following equation, 
\begin{equation}
\phi = V_{\infty} - E_\mathrm{F}
\label{eq:workfunction}
\end{equation}
Work function variation ($\% \phi$) can be expressed as, 
\begin{equation}
\% \phi = \left[ \frac{\phi_f - \phi_i}{\phi_i} \right] \times 100\%,
\label{eq:percent-change in work function}
\end{equation}

Here, $\phi_i$ and $\phi_f$ represent values of the work function before and after gas adsorption. A negative variation of the work function indicates electron emission from the Fermi level to the vacuum level\,\cite{liu2024adsorption}. The values of the work function and their changes due to gas interaction are summarized in Table\,\ref{tab:work_function}. The variation in work function for Cl$_2$ adsorption was the greatest for BNGr-2 and BNGr-4. Pristine BNGr-2 had a work function value of 4.93 eV, and Cl$_2$ interaction increased it to 6.12 eV, positively changing 24.31\%. Other alloys also experienced an increment in work function due to Cl$_2$ exposure, and the percentage changes are sufficient for working as a work function-based sensor. On the other hand, CO and CO$_2$ adsorption altered the work function values of most of the adsorbents to a very limited extent. CO and CO$_2$ adsorption on BNGr-2, BNGr-3, BNGr-4, and BNGr-6 adsorbents resulted in a change in work function of about 6\%-13\% which could be used for a work function-based sensor. BNGr-3 had the potential to be used as a work function-based sensor for NO and NO$_2$, as the percentage changes for these gas molecules were calculated $-$19.32\% and 14.77\%, respectively. Although hBN, BNGr-5, BNGr-8, and graphene systems showed reasonable changes for NO adsorption such as $-$25.11\%, $-$9.09\%, $-$9.25\%, and $-$9.66\%, respectively. BNGr-3 can be a good candidate for NO sensing because of good adsorption properties along with a sufficient work function changing characteristic. NO$_2$ adsorption on BNGr-6 and BNGr-7 also led to high percentage change values such as 26.38\% and 17.34\% respectively. However, BNGr-3 can be considered for NO$_2$ sensing for maintaining both criteria of acceptable adsorption and work function modifying properties. Work function values of HCN/BNGr-2 and BNGr-2 were 5.442 eV and 4.925 eV, respectively, and the change was 10.5\%. Other systems did not produce noticeable variation in the  work function except HCN/BNGr-4 (15.53\%) and HCN/BNGr-6 (14.11\%).

\subsubsection{Conductivity, Chemiresistive Sensitivity, and Selectivity}
Substantial charge transfer during gas adsorption results in a change in the conductivity of the sensor material. Hence, conductivity-based sensitivity demonstrates the extent of chemiresistive sensing at the working temperature. The change in the band gap contributes to the variation in conductivity, given by the following expressions\,\cite{zhao2020gas}, 
\begin{equation}
\sigma = A \exp\left( -\frac{E_g}{2k_B T} \right)
\label{conductivity}
\end{equation}
\begin{equation}
S =
\frac{\delta\sigma}{\sigma}
=
\frac{
\exp\left(-\frac{E_g(\mathrm{gas+layer})}{2k_BT}\right)
-
\exp\left(-\frac{E_g(\mathrm{layer})}{2k_BT}\right)
}{
\exp\left(-\frac{E_g(\mathrm{layer})}{2k_BT}\right)
}
\label{eq:sensitivity}
\end{equation}

where $\sigma$ is the conductivity, $E_g$ is the band gap, $k_B$ is the Boltzmann constant, T is the working temperature, and $E_g(\mathrm{layer}$ and $E_g(\mathrm{gas+layer}$ are the band gap of the system before and after interaction with gas, respectively. Conductivity and chemiresistive sensitivity of gas adsorption on the most favorable structures (BNGr-2 and BNGr-3) are presented in Table\,\ref{tab:sensitivity} for temperatures 300 K, 400 K, and 500 K. The change in band gap due to gas adsorption should modify the conductance of the systems, which would later reveal the sensitivity of the systems. Sensitivity decreased for all gases with temperature as expected. Among the considered gases, CO and CO$_2$ were calculated to be the least sensitive towards BNGr-2 and BNGr-3. CO/BNGr-2, CO/BNGr-3, CO$_2$/BNGr-2, and CO$_2$/BNGr-3 had very low band gap changes such as 0.013 eV, 0.007 eV, 0.006 eV, and 0.003 eV, respectively. Consequently, sensitivity values for these gases were less than 1, indicating a very low sensitivity to BNGr-2 and BNGr-3. HCN adsorption also had minimal influence on the band gaps of BNGr-2 and BNGr-3, as the band gap difference was only 0.015 eV and 0.013 eV, and sensitivity values at 300K were 11.85 and 0.22, respectively. Cl$_2$/BNGr-2 and Cl$_2$/BNGr-3 had a change in band gaps of 0.205 eV and 0.166 eV, respectively, resulting in a sensitivity of 51.72 and 23.79 at 300K. Very high sensitivity of NO and NO$_2$ to BNGr-2 and BNGr-3 ($\sim10^{6}-10^{17}$) resulted from significant band gap changes during adsorption. Band gaps decreased by 2.116 eV for NO/BNGr-2, 1.288 eV for NO/BNGr-3, 1.76 eV for NO$_2$/BNGr-2, and 1.85 eV for NO$_2$/BNGr-3. Therefore, BNGr-2 and BNGr-3 both could be suitable as chemiresistive sensors for selective detection of NO and NO$_2$. 

\begin{table*}
\fontsize{9pt}{12pt}\selectfont
\setlength{\tabcolsep}{1pt}
  \caption{\ Electrical conductivity ($\sigma$), sensitivity (S), and selectivity of gas adsorbed systems at various temperatures (T)}
\label{tab:sensitivity}

\begin{tabular*}{\textwidth}{@{\extracolsep{\fill}}ccccccccc}
\hline

System &
$\sigma$ (S/m) &
$\sigma$ (S/m) &
$\sigma$ (S/m) &
$S$ &
$S$ &
$S$ &
Selectivity \\

& (T = 300 K) & (T = 400 K) & (T = 500 K)
& (T = 300 K) & (T = 400 K) & (T = 500 K) & (T = 300 K) \\

\midrule

BNGr-2
& $4.522 \times 10^{-21}$
& $5.514 \times 10^{-16}$
& $6.211 \times 10^{-13}$
& --
& --
& --
& -- \\

Cl/BNGr-2
& $2.384 \times 10^{-19}$
& $1.079 \times 10^{-14}$
& $6.705 \times 10^{-12}$
& 51.7219
& 18.5656
& 9.7943
& $1.149 \times 10^{16}$ \\

CO/BNGr-2
& $5.814 \times 10^{-21}$
& $6.659 \times 10^{-16}$
& $7.223 \times 10^{-13}$
& 0.2859
& 0.2075
& 0.1628
& $2.080 \times 10^{18}$ \\

CO$_2$/BNGr-2
& $4.026 \times 10^{-21}$
& $5.055 \times 10^{-16}$
& $5.794 \times 10^{-13}$
& 0.1096
& 0.0834
& 0.0673
& $5.428 \times 10^{18}$ \\

NO/BNGr-2
& $2.689 \times 10^{-3}$
& 0.0118
& 0.0287
& $5.947 \times 10^{17}$
& $2.142 \times 10^{13}$
& $4.619 \times 10^{10}$
& 1 \\

NO$_2$/BNGr-2
& $2.749 \times 10^{-6}$
& $6.752 \times 10^{-5}$
& $4.608 \times 10^{-4}$
& $6.080 \times 10^{14}$
& $1.224 \times 10^{11}$
& $7.419 \times 10^{8}$
& 978.1018 \\

HCN/BNGr-2
& $3.946 \times 10^{-20}$
& $2.799 \times 10^{-15}$
& $2.799 \times 10^{-15}$
& 11.8468
& 5.7857
& 3.6268
& $5.020 \times 10^{16}$ \\

\midrule

BNGr-3
& $9.395 \times 10^{-17}$
& $9.543 \times 10^{-13}$
& $2.420 \times 10^{-10}$
& --
& --
& --
& -- \\

Cl/BNGr-3
& $2.330 \times 10^{-15}$
& $1.060 \times 10^{-11}$
& $1.661 \times 10^{-9}$
& 23.7966
& 10.1121
& 5.8649
& $2.771 \times 10^{9}$ \\

CO/BNGr-3
& $8.205 \times 10^{-17}$
& $8.621 \times 10^{-13}$
& $2.231 \times 10^{-10}$
& 0.1266
& 0.0966
& 0.0780
& $5.207 \times 10^{11}$ \\

CO$_2$/BNGr-3
& $8.865 \times 10^{-17}$
& $9.136 \times 10^{-13}$
& $2.337 \times 10^{-10}$
& 0.0564
& 0.0426
& 0.0342
& $1.169 \times 10^{12}$ \\

NO/BNGr-3
& $1.902 \times 10^{-5}$
& $2.880 \times 10^{-4}$
& $1.471 \times 10^{-3}$
& $6.594 \times 10^{10}$
& $1.301 \times 10^{8}$
& $3.101 \times 10^{6}$
& 1 \\

NO$_2$/BNGr-3
& $1.030 \times 10^{-15}$
& $5.766 \times 10^{-12}$
& $1.020 \times 10^{-9}$
& 2.5842
& 1.6049
& 1.1509
& $2.552 \times 10^{10}$ \\

HCN/BNGr-3
& $7.306 \times 10^{-17}$
& $7.903 \times 10^{-13}$
& $2.081 \times 10^{-10}$
& 0.2223
& 0.1719
& 0.1400
& $2.966 \times 10^{11}$ \\

\hline
\end{tabular*}
\end{table*}

A gas sensor should be able to distinguish a target gas molecule from the other gas molecules present near the surface. To rate this property, the selectivity coefficient was calculated using the following expression\,\cite{sharma2024porous}, 

\begin{equation}
K_{\text{sel}} = \frac{S_{\text{t}}}{S_{\text{i}}}
\label{eq:selectivity}
\end{equation}

where $K_{\text{sel}}$, S$_{t}$, and S$_{i}$ are the selectivity coefficient, sensitivity of the highest
sensitive gas towards a specific system, and sensitivity of the
other interfering gases, respectively. Selectivity of gas molecules for adsorption on BNGr-2 and BNGr-3 at 300 K is presented in Table\,\ref{tab:sensitivity}. NO, being the most sensitive gas for both structures, was treated as the target gas during selectivity calculation. It was evident that these adsorbents have distinctive selectivity quantities for most of the gas molecules. However, BNGr-2 showed limited discrimination between CO and CO$_2$ (K$_{sel}\sim10^{18}$), and Cl$_2$ and HCN (K$_{sel}\sim10^{16}$). Despite this, BNGr-2 exhibited a pronounced disparity in selectivity for NO and NO$_2$, indicating ability to distinguish between these two gases. BNGr-3 was observed to have distinct selectivity values for Cl$_2$, CO$_2$, NO, and NO$_2$. However, BNGr-3 demonstrated limited differentiation between CO and HCN (K$_{sel}\sim10^{11}$).

\subsubsection{Reusability}
The reusability of a gas sensor can be estimated from the time required for desorption of the gas molecule from the sensor surface, termed as recovery time. According to the transition-state theory, recovery time ($\tau$) can be defined as:
\begin{equation}
\tau = \nu^{-1} \exp\left(\frac{-E_{ad}}{k_B T}\right)
\label{eq:recovery_time}
\end{equation}

where $\nu$ is the attempt frequency, $k_B$ is the Boltzmann constant, and T is the temperature. A shorter recovery time indicates rapid regeneration of the sensor surface, whereas a longer recovery time suggests a stronger gas-adsorbent interaction and slow regeneration. Recovery time values for gas/BNGr-2 and gas/BNGr-3 at varying temperatures are presented in Table\,\ref{tab:recovery_time}. Cl$_2$/BNGr-2 needed the longest time ($\sim10^{7}$ s) to desorb gas at 300 K, which indicated strong interaction. Such a longer recovery time would enhance adsorption characteristics, although slowed desorption would impede reusability. However, the Cl$_2$/BNGr-3 system had a relatively faster recovery (191 s) at room temperature, demonstrating balanced adsorption-desorption characteristics. CO and CO$_2$ were calculated to be released quickly ($\sim10^{-3}-10^{-4}$ s) from the surface of BNGr-2 and BNGr-3. It was evident from Table\,\ref{tab:recovery_time} that NO and NO$_2$ would be released comparatively faster from BNGr-2 than from BNGr-3 due to the shorter recovery time frame. Recovery of HCN/BNGr-2 and HCN/BNGr-3 was found to be very fast as well, in the range of $\sim10^{-1}-10^{-2}$ s at 300 K. For  higher temperatures as 400 K and 500 K, the calculated recovery times were much shorter, depicting the instant release of the gas molecules. These findings suggest that BNGr-2 and BNGr-3 were competent as reusable gas sensors.

\begin{table*}
\fontsize{9pt}{12pt}\selectfont
\setlength{\tabcolsep}{1pt}
  \caption{\ Recovery time of BNGr-2 and BNGr-3 sensors at different temperatures}
\label{tab:recovery_time}

\begin{tabular*}{\textwidth}{@{\extracolsep{\fill}}l cc cc cc}
\hline
\multirow{2}{*}{Gas} &
\multicolumn{2}{c}{Recovery time at 300 K (s)} &
\multicolumn{2}{c}{Recovery time at 400 K (s)} &
\multicolumn{2}{c}{Recovery time at 500 K (s)} \\
\cmidrule(lr){2-3} \cmidrule(lr){4-5} \cmidrule(lr){6-7}
 & BNGr-2 & BNGr-3 & BNGr-2 & BNGr-3 & BNGr-2 & BNGr-3 \\
\midrule
Cl$_2$ & $1.38\times10^{7}$ & $1.91\times10^{2}$ & $1.81\times10^{3}$ & $4.10\times10^{-1}$ & $8.44$ & $1.03\times10^{-2}$ \\
CO     & $1.89\times10^{-4}$ & $1.46\times10^{-4}$ & $1.29\times10^{-5}$ & $1.06\times10^{-5}$ & $2.56\times10^{-6}$ & $2.19\times10^{-6}$ \\
CO$_2$ & $2.91\times10^{-3}$ & $1.66\times10^{-3}$ & $9.98\times10^{-5}$ & $6.57\times10^{-5}$ & $1.32\times10^{-5}$ & $9.44\times10^{-6}$ \\
NO     & $8.90\times10^{-4}$ & $3.03\times10^{-2}$ & $4.11\times10^{-5}$ & $5.79\times10^{-4}$ & $6.49\times10^{-6}$ & $5.38\times10^{-5}$ \\
NO$_2$ & $1.09\times10^{-2}$ & $8.65\times10^{1}$ & $2.69\times10^{-4}$ & $2.26\times10^{-1}$ & $2.91\times10^{-5}$ & $6.37\times10^{-3}$ \\
HCN    & $1.16\times10^{-1}$ & $2.96\times10^{-2}$ & $1.58\times10^{-3}$ & $5.69\times10^{-4}$ & $1.20\times10^{-4}$ & $5.31\times10^{-5}$ \\
\hline
\end{tabular*}
\end{table*}

\subsubsection{Optical Absorption}
\begin{figure*}
 \centering
 \includegraphics[width=\textwidth]{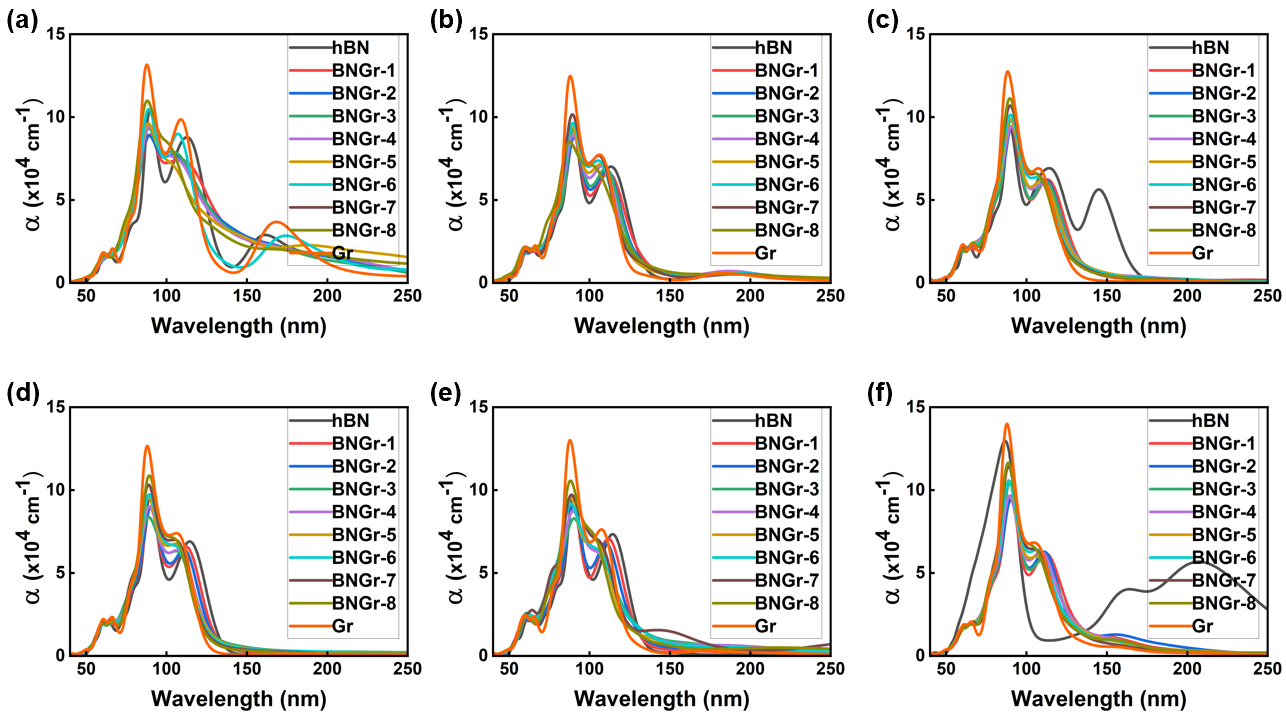}
 \caption{ Absorption coefficient of gas adsorbed systems (a) Cl$_2$/alloys, (b) CO/alloys, (c) CO$_2$/alloys, (d) NO/alloys, (e) NO$_2$/alloys, and (f) HCN/alloys.}
 \label{fig:Absorption_coeff}
\end{figure*}

Optical absorption spectra were calculated for the gas adsorbed systems while incident light was applied in the z direction. The effect of gas exposure on the optical properties of the adsorbents can be observed in Fig\,\ref{fig:Absorption_coeff}. Cl$_2$ adsorbed systems had primary absorption peaks of 8.89$-$13 ($\times10^{4}$) cm$^{-1}$ in the range of 80$-$95 nm wavelength. However, the absorption spectra of Cl$_2$/hBN, Cl$_2$/BNGr-6, and Cl$_2$/graphene showed short secondary peaks near 160-170 nm wavelength. The rest of the systems did not develop any more peaks in their spectra. CO/alloys did not show much disparity in the absorption spectra among themselves. The primary peaks ranged from 8.51$\times10^{4}$ to 12.36$\times10^{4}$ cm$^{-1}$ (wavelength range 83$-$95 nm). Similarly, CO$_2$/alloys systems had closely comparable spectra, except CO$_2$/hBN showed secondary and tertiary peaks at 115 and 145 nm, respectively. Primary peaks of these systems were observed at 9.5$-$12.75 ($\times10^{4}$) cm$^{-1}$ in the range of 83$-$97 nm. NO/alloy systems exhibited primary absorption peaks ranging in the 8.36$-$12.64 ($\times10^{4}$) cm$^{-1}$ region. No secondary peak was observed for alloys that had a carbon ratio larger than 22\%. NO$_2$/alloys exhibited a similar nature in absorption spectra, with primary peaks ranging from 8.29 to 12.98 ($\times10^{4}$) cm$^{-1}$ at 83$-$97 nm wavelength, and no significant secondary peak was observed for gas adsorbed BNGr-3 to graphene systems. HCN/hBN exhibited very distinctive absorption spectra compared to other HCN adsorbed systems in the 150$-$250 nm region. Primary peaks of the HCN/alloy systems ranged from 9.34 to 13.85 ($\times10^{4}$) cm$^{-1}$. It is evident that gases/graphene systems showed the highest absorption peaks. 

\subsubsection{Temperature-Dependent Adsorption}
Gas adsorption is inherently temperature dependent. Previously calculated gas adsorption energies at 0 K from the DFT study would change with varying temperatures. Vibrational energy would contribute to desorbing the gas molecule from the surface. In this study, thermal corrections were obtained by modeling phonon vibrations as quantum harmonic oscillators, and the corresponding vibrational energies were calculated using the finite difference method.  The Helmholtz free energy ($F(T)$) was calculated from the following expression, 

\begin{equation}
F(T) = E_{\mathrm{DFT}} + F_{\mathrm{vib}}(T),
\label{eq:helmholtz}
\end{equation}

where $E_{\mathrm{DFT}}$ is the ground-state electronic energy obtained from density functional theory (DFT), and $F_{\mathrm{vib}}(T)$ is the temperature-dependent vibrational free energy calculated from phonon frequencies using the finite difference method within the harmonic approximation\,\cite{mahmud2024first}. 
\begin{figure*}
 \centering
 \includegraphics[width=\textwidth]{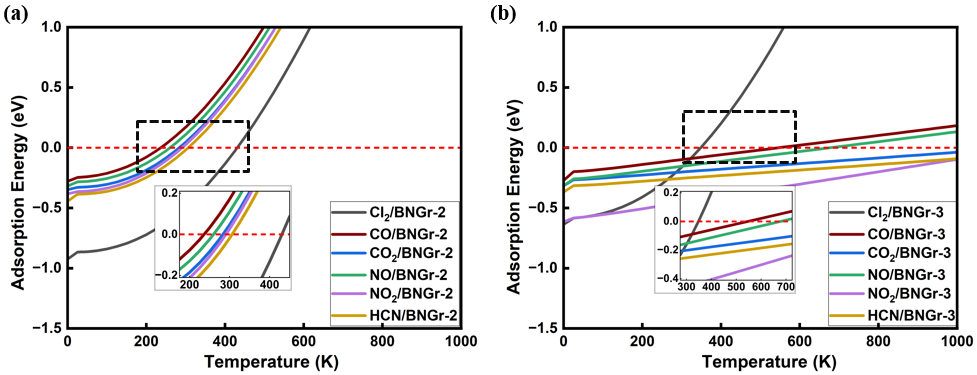}
 \caption{Adsorption energy variation with temperature for gas adsorption on (a) BNGr-2 and (b) BNGr-3.}
 \label{fig:Ads_T}
\end{figure*}
Figure\,\ref{fig:Ads_T} illustrates the variation of adsorption energy with temperature for different gas molecules adsorbed on the BNGr-2 and BNGr-3 alloys. In both systems, the adsorption energy gradually increased with temperature, indicating reduced adsorption stability at elevated temperatures. At lower temperatures, all gases exhibited negative adsorption energies, confirming thermodynamically favorable adsorption on the surfaces. For the BNGr-2 configuration shown in Fig.\,\ref{fig:Ads_T}(a), Cl$_2$ demonstrated the strongest interaction with the surface, exhibiting the most negative adsorption energy among all investigated gases. However, its adsorption energy rose rapidly with temperature and crossed the zero-energy line near 425 K, indicating desorption. The other gas molecules, including CO, CO$_2$, NO, NO$_2$, and HCN, showed endothermic interaction with BNGr-2 at 250$-$300 K. In the BNGr-3 configuration shown in Fig.\,\ref{fig:Ads_T}(b), the adsorption energies displayed a more gradual increase with temperature compared to BNGr-2, indicating enhanced thermal stability of the adsorbed systems. NO$_2$ exhibited the strongest adsorption stability on BNGr-3, maintaining negative adsorption energy over the widest temperature range. In contrast, Cl$_2$ again showed rapid desorption behavior as temperature increased, and the Cl$_2$/BNGr-3 system was spontaneous until 350 K. CO and NO were released from BNGr-2 at 550 K and 700 K, respectively. However, CO$_2$, NO$_2$, and HCN adsorption demonstrated minimal dependence on temperature. Overall, the results suggested that BNGr-2 and BNGr-3 could be promising candidates for temperature controlled gas sensing applications.

\section{Conclusions}
We analyzed the gas-sensing performance of hBN-Gr alloys of different carbon percentages in the presence of toxic gas molecules utilizing DFT-based first-principles study. Carbon substitution contributed to narrowing the band gap of the alloys as new impurity states arrived under the conduction band. The band structures of BNGr-2 and BNGr-3 exhibited band gaps of 2.44 eV and 1.83 eV, respectively. After adding gas molecules, adsorption energies followed a trend such that the larger the carbon fraction, the more adsorption energy was calculated up to an optimum percentage, which was 33\% for NO and NO$_2$, and 22\% for the rest of the gases. Afterward, adsorption energies decreased, indicating that the controlled combination of graphene and hBN outperforms their pristine structures in gas sensing. Strong physical adsorption was observed for Cl$_2$/BNGr-2 and NO$_2$/BNGr-3 systems with adsorption energies of $-$0.93 eV and $-$0.62 eV, respectively. CO/BNGr-2, CO$_2$/BNGr-2, NO/BNGr-3, and HCN/BNGr-2 demonstrated low to moderate physisorption with adsorption energies of $-$0.28 eV, $-$0.35 eV, $-$0.41 eV, and $-$0.44 eV, respectively. Charge transfer and EDD analysis supported the adsorption types and processes quantitatively and graphically. Hirshfeld charge transfer of Cl$_2$/BNGr-2 was $-$0.23 e, and NO$_2$/BNGr-3 was $-$0.21 e, which indicated pronounced interaction. From the TDOS and PDOS plots of the gas-adsorbed systems, significant hybridization was observed between the C and B atoms of the adsorbents and gas molecules. NO and NO$_2$ exhibited the highest sensitivity ($\sim$$10^{14}-10^{17}$ at room temperature). Distinctive selectivity quantities were found for NO and NO$_2$ at 300 K. Work function variation of an average of 7\%$-$20\% indicated the potential of the proposed systems as work function-based sensors. Significantly low recovery times ($\sim$$10^{-1}$-$10^{-4}$ seconds) were found for the CO, CO$_2$, NO, NO$_2$, and HCN adsorbed BNGr-2 and BNGr-3, revealing these structures to be very convenient as reusable gas sensors at room temperature. Distinctive optical absorption behavior was found for the systems within a 70$-$120 nm wavelength range. Furthermore, the temperature-dependent release method revealed that the alloy BNGr-2 is capable of releasing all types of gas molecules within 425 K. Overall, these findings indicated the strong potential of the proposed material for the development of high-performance gas sensing devices.


\section*{Author contributions}
\textbf{Tanjuma Shikder Jhumu:} conceptualization, formal analysis, methodology, visualization, software, investigation, and writing -original draft. \textbf{Ahmed Zubair:} supervision, conceptualization, methodology, visualization, project administration, resources, writing– original draft, and writing – review \& editing

\section*{Conflicts of interest}
There are no conflicts to declare.

\section*{Data availability}
Source data are available from the corresponding author upon request. 


\section*{Acknowledgements}

T.S.J. acknowledges the Nanoscale Simulation, Characterization and Fabrication Lab, Department of EEE, BUET, supervised by A. Z., for this work. All the authors thank the Department of EEE, BUET, for providing the necessary support. 



\balance


\bibliography{rsc} 
\bibliographystyle{rsc} 
\end{document}